\documentclass[epj]{svjour}
\usepackage{latexsym,amssymb}
\usepackage{graphics}
\begin{document}
\title{Thermoelectric Transport Properties in Disordered Systems Near the 
Anderson Transition}
\author{C. Villagonzalo\thanks{\email{villagonzalo@physik.tu-chemnitz.de}} \and R.\,A.\,R\"{o}mer \and M. Schreiber
}                     
%
%
\institute{Institut f\"{u}r Physik, Technische Universit\"{a}t,
  D-09107 Chemnitz, Germany}
%
\date{Received: $Revision: 1.10 $
}
%
\abstract{We study the thermoelectric transport properties in the
  three-dimensional Anderson model of localization near the
  metal-insulator transition (MIT). In particular, we investigate the
  dependence of the thermoelectric power $S$, the thermal conductivity
  $K$, and the Lorenz number $L_0$ on temperature $T$. We first
  calculate the $T$ dependence of the chemical potential $\mu$ from
  the number density $n$ of electrons at the MIT using an averaged
  density of states obtained by diagonalization.  Without any
  additional approximation, we determine from $\mu(T)$ the behavior of
  $S$, $K$ and $L_0$ at low $T$ as the MIT is approached.  We find
  that $\sigma$ and $K$ decrease to zero at the MIT as
  $T\rightarrow{0}$ and show that $S$ does not diverge.  Both $S$ and
  $L_0$ become temperature independent at the MIT and depend only on
  the critical behavior of the conductivity.
} 
\maketitle
\section{Introduction}\label{intro}

The Anderson-type metal-insulator transition (MIT) has been the
subject of investigation for decades since Anderson formulated the
problem in 1958 \cite{anderson}.  He proposed that increasing the
strength of a random potential in a three-dimensional (3D) lattice may
cause an ``absence of diffusion'' for the electrons.  Today, it is
widely accepted that near this exclusively-disorder-induced MIT the
d.\ c.\ conductivity $\sigma$ behaves as $|E-E_c|^\nu$, where $E_c$ is
the critical energy or the mobility edge at which the MIT occurs, and
$\nu$ is a universal critical exponent \cite{kramer}.  Numerical
studies based on the Anderson Hamiltonian of localization have
supported this scenario with much evidence
\cite{kramer,schreiber,bulka,hofstetter,slevin}.  In measurements of
$\sigma$ near the MIT in semiconductors and amorphous alloys this
behavior was also observed with varying values of $\nu$ ranging from
$0.5$--$1.3$ \cite{nu,lauinger,stupp}.  It is currently believed that
these different exponents are caused by interactions in the system
\cite{belitz}. Indeed, an MIT may be induced not only by disorder but
also by interactions such as electron-electron and electron-phonon
interactions, among others \cite{mott2}.  Nevertheless, the
experimental confirmation of the critical behavior of $\sigma$ allows
the use of the Anderson model in order to describe the transition
between the insulating and the metallic states in disordered systems.

Besides for the conductivity $\sigma$, experimental investigations can
also be done for thermoelectric transport properties such as the
thermoelectric power $S$ \cite{lauinger,sherwood,lakner}, the thermal
conductivity $K$ and the Lorenz number $L_0$.  The behavior of these
quantities at low temperature $T$ in disordered systems close to the
MIT has so far not been satisfactorily explained.  In particular, some
authors have argued that $S$ diverges \cite{sherwood,castellani} or
that it remains constant \cite{sivan,enderby} as the MIT is approached
from the metallic side.
In addition, $|S|$ at the MIT has been predicted \cite{enderby} to be
of the order of $\sim 200\,\mu$V/K. On the other hand, measurements of
$S$ close to the MIT conducted on semiconductors for $T\leq1\,$K
\cite{lakner} and on amorphous alloys in the range
$5\,$K$\leq{T}\leq350\,$K \cite{lauinger} yield values of the order of
0.1-1$\,\mu$V/K. They also showed that $S$ can either be negative or
positive depending on the donor concentration in semiconductors or the
chemical composition of the alloy.  The large difference between the
theoretical and experimental values is still not resolved.

The objective of this paper is to study the behavior of the
thermoelectric transport properties for the {\em Anderson model}\/ of
localization in disordered systems near the MIT at low $T$.  We
clarify the above mentioned difference in the theoretical calculations
for $S$, by showing that the radius of convergence for the Sommerfeld
expansion used in Refs.\ \cite{castellani,sivan} is zero at the MIT.
We show that $S$ is a finite constant at the MIT as argued in Refs.\ 
\cite{sivan,enderby}.  Besides for $S$, we also compute the $T$
dependence for $\sigma$, $K$, and $L_0$.  Our approach is neither
restricted to a low- or high-$T$ expansion as in Refs.\ 
\cite{castellani,sivan}, nor confined to the critical regime as in
Ref.\ \cite{enderby}.

We shall first introduce the model in Sec.\,\ref{Amodel}.  Then in
Secs.\,\ref{linear} and \ref{sec-etc} we review the thermoelectric
transport properties in the framework of linear response and the
present formulations in calculating them. In Sec.\,\ref{method} we
shall show how to calculate the $T$ dependence of these properties.
Results of these calculations are then presented in
Sec.\,\ref{result}.  Lastly, in Sec.\,\ref{conclude} we discuss the
relevance of our study to the experiments.

\section{The Anderson Model of Localization}\label{Amodel}
\begin{figure}
\hspace*{0.75cm}
\resizebox{0.4\textwidth}{!}{
  \includegraphics{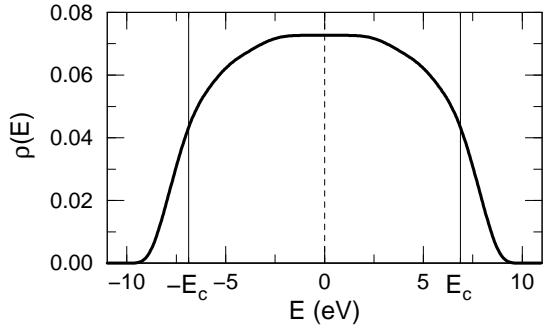}
}
\caption{
  The density of states of a 3D Anderson model, averaged over many
  disorder realizations with $W=12$.  The solid vertical lines at
  $-E_{c}$ and $E_c$ denote the mobility edges.}
\label{dosfig}      
\end{figure}

The Anderson model \cite{anderson} is described by the Hamiltonian
\begin{equation}
H=\sum_{i}\epsilon_{i}|i\rangle\langle{i}|+\sum_{i\neq{j}}t_{ij}
|i\rangle\langle{j}| \label{hamilton}
\end{equation}
where $\epsilon_{i}$ is the potential energy at the site $i$ of a
regular cubic lattice and is assumed to be randomly distributed in the
range $[-W/2,W/2]$ throughout this work.  The hopping parameters
$t_{ij}$ are restricted to nearest neighbors. For this system, at
strong enough disorder and in the absence of a magnetic field, the
one-particle wavefunctions become exponentially localized at $T=0$ and
$\sigma$ vanishes \cite{kramer}.  Illustrating this, we refer to
Fig.\,\ref{dosfig}
where we show the density of states $\rho(E)$ obtained by
diagonalizing the Hamiltonian (\ref{hamilton}) with the Lanczos method
as in Ref.\,\cite{milde0,milde}.  The states in the band tails with energy
$|E|>E_c$ are localized within finite regions of space in the system
at $T=0$ \cite{kramer}.  When the Fermi energy $E_F$ is within these
tails at $T=0$ the system is insulating.  Otherwise, if
$|E_{F}|<E_{c}$ the system is metallic.  The critical behavior of
$\sigma$ is given by
\begin{equation}
\sigma(E)=\left\{ \begin{array}{cc}
\sigma_{0}\left|1-\frac{E}{E_{c}}\right|^{\nu}, & \quad |E| \leq E_{c}, \\
0,                                   & \quad |E|>E_{c},
\end{array}\right. \label{dc_cond}
\end{equation}
where $\sigma_{0}$ is a constant and $\nu$ is the conductivity
exponent \cite{kramer}.  Thus, $E_{c}$ is called the mobility edge
since it separates localized from extended states. At the critical
disorder $W_c=16.5$, the mobility edge occurs at $E_c=0$, all states
with $|E|>0$ are localized \cite{schreiber,bulka} and states with
$E=0$ are multifractal \cite{schreiber,milde0}. The value of $\nu$ has
been computed from the non-linear sigma-model \cite{wegner},
transfer-matrix methods \cite{kramer,slevin}, Green functions methods
\cite{kramer}, and energy-level statistics \cite{hofstetter,els}.
Here we have chosen $\nu= 1.3$, which is in agreement with
experimental results in Si:P \cite{stupp} and the numerical data of
Ref.\ \cite{hofstetter}. More recent numerical results
\cite{kramer,slevin}, computed with higher accuracy, suggest that $\nu
= 1.5\pm 0.1$. As we shall show later, this difference only slightly
modifies our results.
We emphasize that the Hamiltonian (\ref{hamilton}) only incorporates
the electronic degrees of freedom of a disordered system and further
excitations such as lattice vibrations are not included.

For comparison with the experimental results, we measure $\sigma$ in
Eq.\,(\ref{dc_cond}) in units of $\mathrm{\Omega}^{-1}\mbox{cm}^{-1}$.
We fix the energy scale by setting $t_{ij} = 1$ eV. Hence the band
width of Fig.\ \ref{dosfig} is comparable to the band width of
amorphous alloys \cite{haussler}.  Furthermore, the experimental
investigations of the thermoelectric power $S$ in amorphous alloys
\cite{lauinger} have been done at high electron filling \cite{private}
and thus we will mostly concentrate on the MIT at $E_c$.

\section{Linear Thermoelectric Effects}\label{linear}

\subsection{Definition of the Transport Properties}
\begin{figure}
\hspace*{0.75cm}
\resizebox{0.4\textwidth}{!}{
  \includegraphics{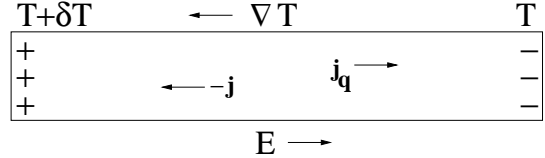}
}
\caption{
  In an open circuit, a temperature gradient $\nabla{T}$ induces an
  electric field $\mathbf{E}$ in the opposite direction which opposes
  the thermal flow of electrons.}
\label{bar}       
\end{figure}

Thermoelectric effects in a system are due mainly to the presence of a
temperature gradient $\mathbf{\nabla}T$ and an electric field
$\mathbf{E}$ \cite{ashcroft}.  We recall that in the absence of
$\mathbf{\nabla}T$ with $\mathbf{E}\neq{0}$, the electric current
density $\langle\mathbf{j}\rangle$ flowing at a point in a conductor
is directly proportional to $\mathbf{E}$,
\begin{equation}
\langle\mathbf{j}\rangle=\sigma\mathbf{E}\;. \label{sigma1}
\end{equation}
By applying a finite gradient $\mathbf{\nabla}T$ in an open circuit,
electrons, the thermal conductors, would flow towards the low-$T$ end
as shown in Fig.\,\ref{bar}. This causes a build-up of negative
charges at the low-$T$ end and a depletion of negative charges at the
high-$T$ end. Consequently, this sets up an electric field
$\mathbf{E}$ which opposes the thermal flow of electrons.  For small
$\mathbf{\nabla}T$, it is given as
\begin{equation}
\mathbf{E}=S\mathbf{\nabla}{T}\,. \label{tp1}
\end{equation}
This equation defines the \textit{thermopower} $S$.  In the Sommerfeld
free electron model of metals, $S$ is found to be directly
proportional to $-T$ \cite{ashcroft}. Note that the negative sign is
brought about by the charge of the thermal conductors.
For small $\mathbf{\nabla}T$, the flow of heat in a system is
proportional to $\nabla{T}$. Fourier's Law gives this as
\begin{equation}
\langle{\mathbf{j}_{q}}\rangle=K(-\mathbf{\nabla}T) \label{tc1}
\end{equation}
where $\langle{\mathbf{j}_{q}}\rangle$ is the heat current density and
$K$ is the thermal conductivity \cite{ashcroft}. At low $T$, the
phonon contribution to $\sigma$ and $K$ becomes negligible compared to
the electronic part \cite{ashcroft}.  As $T\rightarrow{0}$, $\sigma$
approaches a constant and $K$ becomes linear in $T$.  One can then
verify the empirical law of Wiedemann and Franz which says that the
ratio of $K$ and $\sigma$ is directly proportional to $T$
\cite{wiedemann,chester}.  The proportionality coefficient is
known as the Lorenz number $L_0$,
\begin{equation}
L_{0}=\frac{e^2}{k_{B}^2}\frac{K}{\sigma{T}}\, \label{lo1}
\end{equation}
where $e$ is the electron charge and $k_{B}$ is the Boltzmann
constant.  For metals, it takes the universal value $\pi^{2}/3$
\cite{ashcroft,chester}.  Strictly speaking, the law of Wiedemann and
Franz is valid at very low $T$ ($\lesssim{10}\,$K) and at high (room)
$T$.  This is because in these regions the electrons are scattered
elastically. At $T\sim10-100\,$K deviations from the law are observed
which imply that $K/\sigma{T}$ depends on $T$.

In summary, Eqs.\,(\ref{sigma1})-(\ref{lo1}) express the
phenomenological description of the transport properties.
 
\subsection{The Equations of Linear Response}
A more compact and general way of looking at these thermoelectric
``forces'' and effects is as follows: the responses of a system to
$\mathbf{E}$
and $\mathbf{\nabla}T$ up to linear order \cite{callen} are
\begin{equation}
\langle\mathbf{j}\rangle = |e|^{-1}\left(|e|L_{11}
\mathbf{E}-L_{12}T^{-1}\mathbf{\nabla}T\right) \label{ecurrent}
\end{equation}
and
\begin{equation}
\langle\mathbf{j}_{q}\rangle = |e|^{-2}\left(|e|L_{21}
\mathbf{E}-L_{22}T^{-1}\mathbf{\nabla}T\right) \label{hcurrent}.
\end{equation}
The kinetic coefficients $L_{ij}$ are the keys to calculating the
transport properties theoretically.  Using Ohm's law (\ref{sigma1}) in
Eq.\,(\ref{ecurrent}), we obtain
\begin{equation}
\sigma = L_{11}\,. \label{sigma2} 
\end{equation}
Also from Eq.\,(\ref{ecurrent}), $S$, measured under the condition of
zero electric current, is expressed as
\begin{equation}
S = \frac{L_{12}}{|e|TL_{11}}\;. \label{tp2}
\end{equation}
With the same condition,
Eq.\,(\ref{hcurrent}) yields
\begin{equation}
K=\frac{L_{22}L_{11}-L_{21}L_{12}}{|e|^{2}TL_{11}}\,. \label{tc2}
\end{equation}
From Eq.\ (\ref{lo1}) $L_0$ is given as
\begin{equation}
L_{0}=\frac{L_{22}L_{11}-L_{21}L_{12}}{(k_{B}TL_{11})^2}\,. \label{lo2}
\end{equation}
Therefore, we will be able to determine the transport properties once
we know the coefficients $L_{ij}$. We note that in the absence of a
magnetic field, as considered in this work, the Onsager
relation $L_{21}=L_{12}$ holds \cite{callen}.

Eliminating the kinetic coefficients in Eqs.\,(\ref{ecurrent}) and
(\ref{hcurrent}) in favor of the transport properties, we obtain
\begin{equation}
  \langle\mathbf{j}\rangle =
  \sigma\mathbf{E}-\sigma{S}\nabla{T}\label{ecurrent2}
\end{equation}
and
\begin{equation}
  \frac{\langle\mathbf{j}_{q}\rangle}{T} = S\langle\mathbf{j}\rangle
  -\frac{K\nabla{T}}{T}.\label{entropy}
\end{equation}
Here, $\langle\mathbf{j}_{q}\rangle/T$ is simply the entropy current
density \cite{callen}.  Hence, the thermopower is just the entropy
transported per Coulomb by the flow of thermal conductors.  According
to the third law of thermodynamics, the entropy of a system and, thus,
also $\langle\mathbf{j}_{q}\rangle/T$ will go to zero as
$T\rightarrow{0}$.  We can  check with Eqs.\,(\ref{ecurrent2}) and
(\ref{entropy}) that this is satisfied by our calculations in the 3D
Anderson model.

\subsection{Application to the Anderson Transition}
In general, the linear response coefficients $L_{ij}$ are obtained
through the Chester-Thellung-Kubo-Greenwood (CTKG) formulation
\cite{chester,kubo}. The kinetic coefficients are expressed as
\begin{equation}
L_{11}=  \int_{-\infty}^{\infty} A(E)   
\left[ - \frac{\partial f(E,\mu,T)}{\partial E} \right] dE\,, \label{l11}
\end{equation}
\begin{equation}
L_{12}= - \int_{-\infty}^{\infty} A(E)\left[E-\mu(T)\right] 
\left[- \frac{\partial f(E,\mu,T)}{\partial E} \right] dE\,, \label{l12}
\end{equation}
and
\begin{equation}
L_{22}=  \int_{-\infty}^{\infty} A(E)\left[E-\mu(T)\right]^2
\left[- \frac{\partial f(E,\mu,T)}{\partial E} \right] dE\,, \label{l22}
\end{equation}
where $A(E)$ contains all the system-dependent features, $\mu(T)$ is
the chemical potential and
\begin{equation}
  f(E,\mu,T)=1/\left\{1+\exp([E-\mu(T)]/k_{B}T)\right\}
\label{eq-fermi}
\end{equation}
is the Fermi function.  The CTKG approach inherently assumes that the
electrons are noninteracting and that they are scattered elastically
by static impurities or by lattice vibrations.  A nice feature of this
formulation is that all microscopic details of the system such as the
dependence on the strength of the disorder enter only in $A(E)$.  This
function $A(E)$ can be calculated in the context of the
relaxation-time approximation \cite{ashcroft}.  However, an exact
evaluation of $L_{ij}$ is difficult, if not impossible, since it
relies on the exact knowledge of the energy and $T$ dependence of the
relaxation time. In most instances, these are not known.

In order to incorporate the Anderson model and the MIT in the CTKG
formulation, a different approach is taken: We have seen 
in Eq.\,(\ref{sigma2}) that the d.c.\ conductivity is just $L_{11}$.
Thus, to take into account the MIT in this formulation, we identify
$A(E)$ with $\sigma(E)$ given in Eq.\,(\ref{dc_cond}).
The $L_{ij}$ in Eqs.\,(\ref{l11})-(\ref{l22}) can now be easily
evaluated close to the MIT without any approximation, once the $T$
dependence of the chemical potential $\mu$ is known.  Unfortunately,
this is not known for the experimental systems under consideration
\cite{nu,lauinger,stupp,sherwood,lakner}, nor for the 3D Anderson
model. Thus one has to resort to approximate estimations of $\mu$, as
we do next, or to numerical calculations, as we shall do in the next
sections.

\section{Evaluation of the Transport Coefficients}
\label{sec-etc}

\subsection{Sommerfeld expansion in the metallic regime}
\label{sec-sommerfeld}

Circumventing the computation of $\mu(T)$, one can use that
$-\partial{f}/\partial{E}$ is appreciable only in an energy range of
the order of $k_{B}T$ near $\mu\approx{E_F}$. The lowest non-zero $T$
corrections for the $L_{ij}$ are then accessible by the Sommerfeld
expansion \cite{ashcroft}, provided that $A(E)$ is nonsingular and
slowly varying in this region.  Hence, in the limit $T\rightarrow{0}$,
the transport properties are \cite{sommer}
\begin{equation}
  \sigma=A(E_F)+\frac{\pi^2}{6}(k_{B}T)^{2}\left.\frac{d^{2}A(E)}
    {dE^2}\right|_ {E=E_F}\,,\label{sigma3}
\end{equation}
\begin{equation}
  S=- \frac{\pi^{2}k_{B}^{2}T}{3|e| A(E_{F})}
  \left.\frac{dA(E)}{dE}\right|_{E=E_F}\,, \label{tp3}
\end{equation}
\begin{equation}
  K=\frac{\pi^{2}k_{B}^{2}T}{3e^2}\left\{A(E_F)
    -\frac{\pi^{2}(k_{B}T)^2}{3A(E_F)}
    \left[\frac{dA(E)}{dE}\right]_{E=E_F}^2 \right\}\,,
  \label{tc3}
\end{equation}
and consequently
\begin{equation}
  L_0=\frac{\pi^{2}}{3}\left\{1-\frac{\pi^{2}(k_{B}T)^2}{3[A(E_F)]^2}
    \left[\frac{dA(E)}{dE}\right]_{E=E_F}^2 \right\}. \label{lo3}
\end{equation}
In the derivations of $S$, $K$, and $L_0$, the term of order $T^2$ in
Eq.\ (\ref{sigma3}) has been ignored as is customary. We remark that
the terms of order $T^2$ in Eqs.\,(\ref{tc3}) and (\ref{lo3}) are
usually dropped, too.  In this case in the metallic regime, $L_0$
reduces to the universal value $\pi^2/3$ \cite{ashcroft}.

The above approach was adopted in Refs.\,\cite{castellani} and
\cite{sivan} to study thermoelectric transport properties in the
metallic regime close to the MIT.  From Eq.\ (\ref{tp3}), the authors
deduce
\begin{equation}
S=-\frac{\nu\pi^{2}k_{B}^{2}T}{3|e|(E_{F}-E_{c})}\,. \label{tpCAS}
\end{equation}
In the metallic regime, this linear $T$ dependence of $S$ agrees with
that of the Sommerfeld model of metals \cite{ashcroft}.  However,
setting $A(E) = \sigma(E)$ at the MIT \cite{castellani} in
Eq.\,(\ref{dc_cond}) is in contradiction to the basic assumption of
the Sommerfeld expansion, since it is not smoothly varying at
$E_{F}=E_c$. Thus identifying $A(E) = \sigma(E)$ in Eqs.\ \ref{sigma3}
- \ref{lo3} is only valid in the metallic regime with $k_B T \ll |E_c
- E_F|$.

\subsection{Exact calculation at $\mu(T)= E_c$}
\label{sec-crit}

A different approach taken by Enderby and Barnes is to fix $\mu=-E_c$
at finite $T$ and later take the limit $T\rightarrow{0}$
\cite{enderby}.  Thus, again without knowing the explicit $T$
dependence of $\mu$, the coefficients $L_{ij}$ can be evaluated at the
MIT.  For the transport properties they obtain,
\begin{equation}
  \sigma=\frac{\sigma_{o}\nu (k_{B}T)^{\nu}I_{\nu}}{\left|E_{c}\right|^{\nu}}\,, 
  \label{sigmaEB}
\end{equation}
\begin{equation}
  S = - \frac{k_{B}}{|e|} \frac{\nu+1}{\nu}
  \frac{I_{\nu+1}}{I_{\nu}}\,,
\label{tpEB}
\end{equation}
\begin{equation}
  K=\frac{\sigma_{o}
    (k_{B}T)^{\nu+2}}{e^{2}T\left|E_{c}\right|^{\nu}}\left[(\nu+2)I_{\nu+2}
    -\frac{(\nu+1)^2I_{\nu+1}^2}{\nu{I}_{\nu}}\right]\,, \label{kEB}
\end{equation}
and
\begin{equation}
  L_{0}=\left[\frac{(\nu+2)I_{\nu+2}}{\nu{I}_{\nu}}-
    \frac{(\nu+1)^{2}I_{\nu+1}^2}{(\nu{I}_{\nu})^2}\right].
  \label{loEB}
\end{equation}
Here $I_{1}=\ln2$, $I_{\nu}= (1-2^{1-\nu})\Gamma(\nu)\zeta(\nu)$ for
$\mathrm{Re}(\nu)>0,\;\nu\neq1$, with $\Gamma(\nu)$ and $\zeta(\nu)$
the usual gamma and Riemann zeta functions. 
We see that at the MIT, $S$ does not diverge nor go to zero but
remains a universal constant. Its value depends only on the
conductivity exponent $\nu$.  This is in contrast to the result
(\ref{tpCAS}) of the Sommerfeld expansion.
In addition, we find that $\sigma\propto T^{\nu}$ and $K\propto
T^{\nu+1}$ as $T\rightarrow{0}$.  Hence, $\sigma$ and $K/T$ approach
zero in the same way.  This signifies that the Wiedemann and Franz law
is also valid at the MIT recovering an earlier result in Ref.\ 
\cite{strinati} obtained via diagrammatic methods.  However, at the
MIT, $L_0$ does not approach $\pi^2/3$ but again depends on $\nu$.  We
emphasize that Eqs.\ (\ref{sigmaEB})-(\ref{loEB}) are exact at $T$
values such that $\mu(T)-E_c=0$ \cite{enderby}. Thus the $T$
dependence of $\sigma$, $S$, $K$, and $L_0$ for a given electron
density can only be determined if one knows the corresponding
$\mu(T)$.

\subsection{High-temperature expansion}

In this section, we will study the lowest-order corrections to the
results obtained before with $\mu(T) = E_c$. We do this by expanding
the Fermi function (\ref{eq-fermi}) for $|E_c - \mu(T)| \ll k_B T$.
In addition, we assume $\mu(T) \approx E_F$ for the temperature range
considered.  This procedure gives
\begin{equation}
\hspace*{-0.15cm}
  \sigma = L_{11} = 
  \frac{\sigma_{o}\nu (k_{B}T)^{\nu}}{\left|E_{c}\right|^{\nu}} 
   \left[ 
    I_{\nu}-(\nu-1)I_{\nu-1}\frac{E_c - E_F}{k_B T}
    \right]
\label{eq-ht-sigma}.
\end{equation}
For the thermo\-power, the leading-order correction can be obtained
without expanding $f(E,\mu,T)$ in $L_{11}$ and $L_{12}$. This yields a
constant for $S$ at the MIT \cite{sivan}. We obtain
\begin{equation}
S = - \frac{k_{B}}{|e|} 
 \left[ 
  \frac{\nu+1}{\nu} \frac{I_{\nu+1}}{I_{\nu}}
  + \frac{E_c - E_F}{k_B T}
 \right]\,.
\label{eq-ht-tp}
\end{equation}
For $K$ and $L_0$, we again have to use the expansion of $f(E,\mu,T)$
as in (\ref{eq-ht-sigma}) in order to get non-trivial terms.  The
resulting expressions are cumbersome and we thus refrain from showing
them here. We remark that the basic ingredients used in the high-$T$
expansion are somewhat contradictory, namely, the expansion is valid
for high $T$ such that $|E_c - E_F| \ll k_B T$, whereas $\mu(T)=E_F$
is true only for $T=0$.

At present, we thus have various methods of circumventing the explicit
computation of $\mu(T)$. However, their ranges of validity are not
overlapping and it is a priori not clear whether the assumptions for
$\mu(T)$ are justified for $S$ or any of the other transport
properties close to the MIT.
In order to clarify the situation, we numerically compute $\mu(T)$ in
the next section and then use the CTKG formulation to compute the
thermal properties without any approximation.

\section{The Numerical Method}\label{method}

In Eqs.\,(\ref{l11})-(\ref{l22}), the explicit $T$ dependence of the
coefficients $L_{ij}$ occurs in $f(E,\mu,T)$ and $\mu(T)$. More
precisely, knowing $\mu(T)$, it is straightforward to evaluate the
$L_{ij}$.  We recall that, for any set of noninteracting particles,
the number density of particles $n$ can be determined as
\begin{equation}
n(\mu,T) = \int_{-\infty}^{\infty}dE \rho(E) f(E,\mu,T)\;, \label{numden}
\end{equation}
where $\rho(E)$ is again the density of energy levels (in the unit
volume) as in Fig.\ \ref{dosfig}.  Vice versa, if we know $n$ and
$\rho(E)$ we can solve Eq.\ (\ref{numden}) for $\mu(T)$.  The density
of states $\rho(E)$ for the 3D Anderson model has been obtained for
different disorder strengths $W$ as outlined in Sec.\ \ref{Amodel}. We
determine $\rho(E)$ with an energy resolution of at least $0.1$ meV
($\sim 1$ K).
Using $\rho(E)$, we first numerically calculate $n$ at $T=0$ for the
metallic, critical and insulating regimes using the respective Fermi
energies $|E_F| < E_c$, $E_F = E_c$, and $|E_F| > E_c$.  With
$\mu=E_{F}$, we have
\begin{equation}
n(E_{F})=\int_{-\infty}^{E_F}dE\rho(E)\,.\label{numEF}
\end{equation} 
Next, keeping $n$ fixed at $n(E_F)$, we numerically determine $\mu(T)$
for small $T >0$ such that $|n(E_F) - n(\mu, T)|$ is zero. Then we
increase $T$ and record the respective changes in $\mu(T)$.  Using
this result in Eqs.\ (\ref{l11})--(\ref{l22}) in the CTKG formulation,
we compute $L_{ij}$ by numerical integration and subsequently
determine the $T$ dependent transport properties
(\ref{sigma2})--(\ref{lo2}).

We consider the disorders $W=8$, $12$, and $14$ where we do not have
large fluctuations in the density of states.  These values are not too
close to the critical disorder $W_c$, so that we could clearly observe
the MIT of Eq.\ (\ref{dc_cond}).  The respective values of $E_c$ have
been calculated previously \cite{schreiber} to be close to $7.0$,
$7.5$, and $8.0$. Within our approach, we choose $E_c$ to be equal to
these values.

\section{Results and Discussions}
\label{result}

Here we show the results obtained for $W=12$ with $E_c=7.5$.  The
results for $\sigma$, $K$, and $L_0$ are the same at $-E_c$ and $E_c$
since they are functions of $L_{11}$, $L_{22}$ and $L_{12}^{2}$, only.
On the other hand, this is not true for $S$.

\subsection{The Chemical Potential}

\begin{figure}
\hspace*{0.75cm}
\resizebox{0.4\textwidth}{!}{%
  \includegraphics{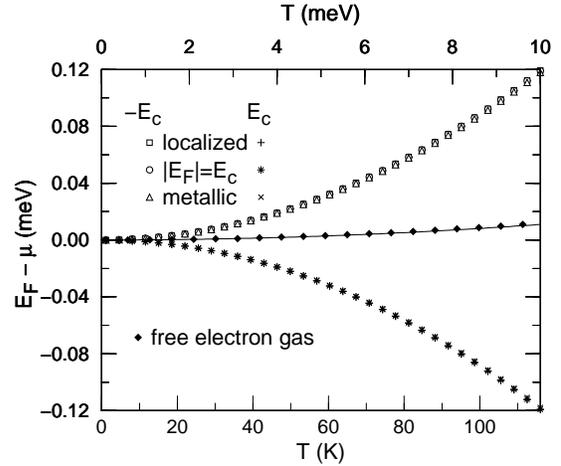}
}                 
\caption{
  The temperature dependence of the chemical potential $\mu$ measured
  with respect to the Fermi energy near both mobility edges.  Also
  shown is $\mu(T)$ for a free electron gas. The solid line denotes
  $\mu(T)$ of Eq.\ (\protect\ref{fegmu}).}
\label{mufig}      
\end{figure}

In Fig.\,\ref{mufig}, we show how $\mu(T)$ behaves for the 3D Anderson
model at $E_F - E_c = 0$, and $\pm 0.01$.  To compare results from
different energy regions we plot the difference of $\mu(T)$ from
$E_F$. We find that $\mu(T)$ behaves similarly in the metallic and
insulating regions and at the MIT for both mobility edges at low $T$.
In all cases we observe $\mu(T)\propto{T}^2$. Furthermore, we see that
$\mu(T)$ at $-E_{c}$ equals $-\mu(T)$ at $E_c$.  This symmetric
behavior with respect to $E_{F}=\mu$ reflects the symmetry of the
density of states at $E=0$ as shown in Fig.\,\ref{dosfig}.  

For comparison and as a check to our numerics, we also compute with
our method $\mu(T)$ of a free electron gas. The density of states is
\cite{ashcroft}
\begin{equation}
\rho(E)= \frac{3}{2} \frac{n}{E_F}\left( \frac{E}{E_F} \right)^{1/2}
\end{equation}
and we again use $E_F=E_c = 7.5$.  We remark that this value of the
mobility edge is in a region where $\rho(E)$ increases with $E$ in an
analogous way as $\rho(E)$ for the Anderson model at $-E_c$ . Thus, as
shown in Fig.\ \ref{mufig}, $\mu(T)$ of a free electron gas is concave
upwards as in the case of the Anderson model at $-E_c$.  We also plot
the result for $\mu(T)$ obtained by the usual Sommerfeld expansion for
Eq.\,(\ref{numden}),
\begin{equation}
  E_{F}-\mu(T)=\frac{E_F}{3}\left(\frac{\pi k_B T}{2E_F}\right)^2.
\label{fegmu}
\end{equation}
We see that our numerical approach is in perfect agreement with the
free electron result.

\subsection{The d.c.\ Conductivity}
\label{dcc}

\begin{figure}
\hspace*{0.75cm}
\resizebox{0.4\textwidth}{!}{
  \includegraphics{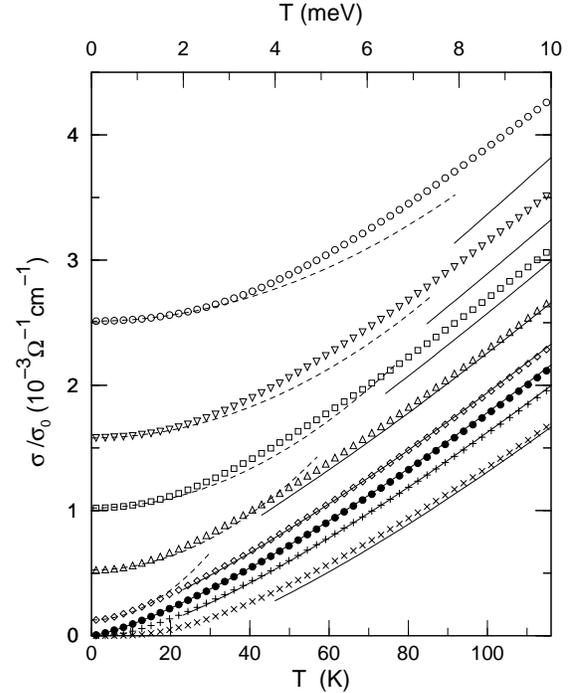}}
\caption{
  The low temperature behavior of the d.c. conductivity $\sigma$. The
  symbols are as shown in Tab.\ \protect\ref{tab:sym}.
  The dashed lines represent the Sommerfeld expansion result for
  $\sigma(T)$ as given in Eq.\,(\protect\ref{sigma3}). For all $8$
  choices of $E_F-E_c$, the corresponding high-$T$ expansion
  (\protect\ref{eq-ht-sigma}) is indicated by solid lines.}
\label{dccondfig}     
\end{figure}

In Fig.\,\ref{dccondfig} we show the $T$ dependence of $\sigma$. The
values of $E_F$ we consider and the corresponding fillings $n$ are
given in Tab.\ \ref{tab:sym}.
\begin{table}[b]
\caption{
  Differences of $E_F$ and $n(E_F)$ with respect to the mobility edge
  at $E_c=7.5$. The density at $E_c$ corresponds to $n= 97.768\%$.}
\label{tab:sym}       
\begin{center}
\begin{tabular}{lr@{}lr@{}lc}
\hline\noalign{\smallskip}
regime & 
\multicolumn{2}{c}{$E_F-E_c$} & 
\multicolumn{2}{c}{$n(E_F)-n(E_c)$} & 
symbol  \\
& 
\multicolumn{2}{c}{(eV)} & 
\multicolumn{2}{c}{($\%$)} & 
\\
\noalign{\smallskip}\hline\noalign{\smallskip}
metallic   & -0. &010       & -0. &031      & $\circ$ \\
           & -0. &007       & -0. &022      & $\bigtriangledown$ \\
           & -0. &005       & -0. &015      & $\Box$ \\
           & -0. &003       & -0. &009      & $\bigtriangleup$ \\
           & -0. &001       & -0. &003      & $\Diamond$ \\
\noalign{\smallskip}
critical   &  0. &000       &  0. &000      & $\bullet$ \\
\noalign{\smallskip}
insulating &  0. &001       &  0. &003      & $+$ \\
           &  0. &003       &  0. &009      & $\times$ \\
           &  0. &010       &  0. &031      & $*$ \\
\noalign{\smallskip}\hline
\end{tabular}
\end{center}
\end{table}
The conductivity at $T=0$ remains finite in the metallic regime with
$\sigma/\sigma_{o}= {|1 - E_F/E_c|}^{\nu}$, because $(-\partial
f/\partial E) \rightarrow \delta(E-E_F)$ in Eq.\ (\ref{l11}) as
$T\rightarrow{0}$.  Correspondingly, we find $\sigma=0$ in the
insulating regime at $T=0$.
In the critical regime, $\sigma(T\rightarrow{0})\sim {T}^{\nu}$, as
derived in Ref.\ \cite{enderby}, see Eq.\,(\ref{sigmaEB}).  We note
that as one moves away from the critical regime towards the metallic
regime one finds within the accuracy of our data that
$\sigma\sim{T}^2$.  We observe that in the metallic regime $\sigma$
increases for increasing $T$.  This is different from the behavior in
a real metal where $\sigma$ decreases with increasing $T$.  However,
as explained in Sec.\ \ref{Amodel}, the behavior of $\sigma$ in
Fig.\,\ref{dccondfig} is due to the absence of phonons in the present
model.

We also show in Fig.\,\ref{dccondfig} results of the Sommerfeld
expansion (\ref{sigma3}) and the high-$T$ expansion
(\ref{eq-ht-sigma}) for $\sigma$.  Paradigmatic for what is to follow
we see that the radius of convergence of the Sommerfeld expansion
decreases for $E_F\rightarrow{E_c}$ and in fact is zero in the
critical regime.  On the other hand, the high-$T$ expansion is very
good in the critical regime down to $T=0$ at $E_c=E_F$. The small
systematic differences between our numerical results and the high-$T$
expansion for large $T$ are due to the differences in $\mu(T)$ and
$E_F$. The expansion becomes worse both in the metallic and insulating
regimes for larger $T$. All of this is in complete agreement with the
discussion of the expansions in Sec.~\ref{sec-etc}.

\subsection{The Thermopower}
\label{tp}

\begin{figure}
\hspace*{0.75cm}
\resizebox{0.4\textwidth}{!}{
  \includegraphics{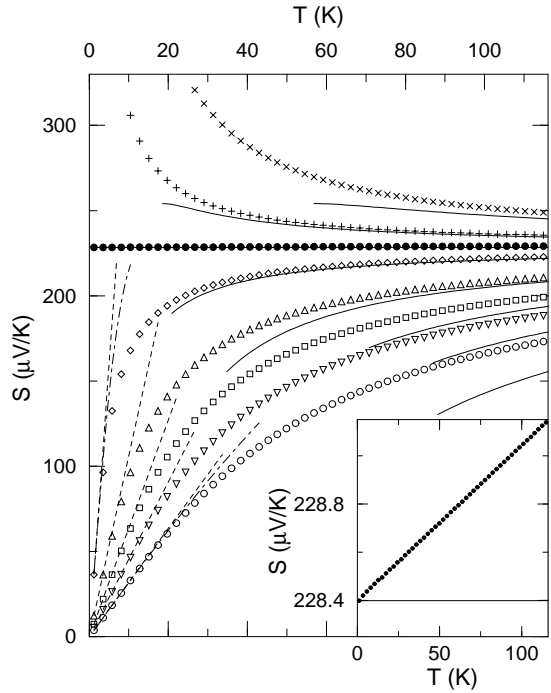}}
\caption{
  The low temperature behavior of the thermopower $S$. The symbols are
  as shown in Tab.\ \protect\ref{tab:sym}.
  The dashed lines represent the behavior of $S(T)$ in the metallic
  regime as given in Eq.\,(\protect\ref{tpCAS}).  The dot-dashed lines
  indicate $S$, calculated with the ${\rm O}(T^2)$ term of Eq.\ 
  (\protect\ref{sigma3}), for $E_F-E_c= -0.01$ eV $(\circ)$ and
  $-0.001$ eV $(\diamond)$. Solid lines are obtained from the high-$T$ 
  expansion (\protect\ref{eq-ht-tp}).  The inset
  shows the behavior at $E_F=E_c$ on an enlarged scale. }
\label{tpfig1}       
\end{figure}

In Fig.\,\ref{tpfig1}, we show the behavior of the thermopower at low
$T$ near the MIT. In the metallic regime, we find $S\rightarrow{0}$ as
$T\rightarrow{0}$.  At very low $T$, $S\propto T$ as predicted by the
Sommerfeld expansion (\ref{tpCAS}).
We see that the Sommerfeld expansion is valid for not too large values
of $T$.  But upon approaching the critical regime, the expansion
becomes unreliable similar to the case of the d.c.\ conductivity of
Sec.\ \ref{dcc}. This behavior persists even if we include higher
order terms in the derivation of $S$ such as the ${\rm O}(T^2)$ term
of Eq.\ (\protect\ref{sigma3}) as shown in Fig.\,\ref{tpfig1}.

Before discussing the critical regime in detail, let us turn our
attention to the insulating regime. Here, $S$ becomes very large as
$T\rightarrow{0}$.  We have observed that it even appears to approach
infinity.  A seemingly divergent behavior in the insulating regime has
also been observed for Si:P \cite{liu}, where it has been attributed
to the thermal activation of charge carriers from $E_F$ to the
mobility edge $E_c$. However, there is a simpler way of looking at
this phenomenon.  We refer again to the open circuit in
Fig.\,\ref{bar}.  Suppose we adjust $T$ at the cooler end such that
$\nabla{T}$ remains constant. As $T\rightarrow{0}$ both $\sigma$ and
$K$ vanish in the case of insulators --- for $K$ we show this in the
next section.  This implies that as $T$ decreases it becomes
increasingly difficult to move a charge from $T$ to
${T}+\delta{T}$. We would need to exert a larger amount of force, and
hence, a larger $\mathbf{E}$ to do the job. From Eq.\,(\ref{tp1}),
this implies a larger $S$ value.

In the critical regime, i.e., setting $E_F = E_c$, we observe in
Fig.\,\ref{tpfig1} that for $T\rightarrow 0$ the thermopower $S$
approaches a value of $228.4\,\mu$V/K. This is exactly the magnitude
predicted \cite{enderby} by Eq.\,(\ref{tpEB}) for $\nu=1.3$. In the
inset of Fig.\,\ref{tpfig1}, we show that the $T$ dependence of $S$ is
linear.  The nondivergent behavior of $S$ clearly separates the
metallic from the insulating regime.  Furthermore, just as for
$\sigma$, the Sommerfeld expansion for $S$ breaks down at
$E_{F}={E_c}$, i.e., the radius of convergence is zero.  Thus, the
divergence of Eq.\,(\ref{tpCAS}) at $E_{F}={E_c}$ reflects this
breakdown and is not physically relevant.  On the other hand, the
high-$T$ expansion \cite{sivan} nicely reflects the behavior of $S$
close to the critical regime as also shown in Fig.\ \ref{tpfig1}. For
$E_F = E_c$, the high-$T$ expansion (\ref{eq-ht-tp}) assumes a
constant value of $S$ for all $T$ due to setting $\mu(T)=E_F$. This is
approximately valid, the differences are fairly small as shown in the
inset of Fig.\,\ref{tpfig1}.

\begin{figure}
\hspace*{0.75cm}
\resizebox{0.4\textwidth}{!}{
  \includegraphics{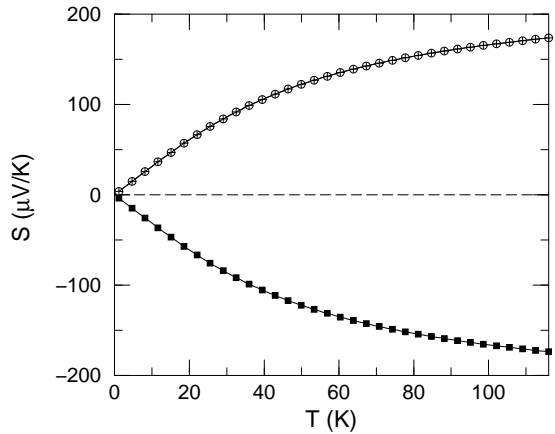}}
\caption{
  An example that the magnitude of $S(T)$ is the same in metallic
  regions close to $-E_c$ ($\blacksquare$) and $E_c$ ($\circ$). The
  $+$-symbols indicate $|S|$ for $-E_c$ and $|E_F-E_c|= 0.01$ eV in all
  cases.}
\label{pnfig}      
\end{figure}

We stress that there is no contradiction that $S>0$ in our
calculations whereas $S<0$ in Ref.\,\cite{enderby}.  In
Fig.\,\ref{pnfig}, we compare $S$ in energy regions close to $E_{c}$
and to $-E_{c}$ \cite{villa}.  Clearly, they have the same magnitude
but $S<0$ at $-E_c$ and $S>0$ at $E_c$. The two cases mainly differ in
their number density $n$. At $-E_c$ the system is at low filling with
$n=2.26\%$ while at $E_c$ the system is at high filling with
$n=97.74\%$.  The sign of $S$ implies that at low filling the
thermoelectric conduction is due to electrons and we obtain the usual
picture as in Fig.\,\ref{bar} where the induced field $\mathbf{E}$ is
in the direction opposite to that of $\nabla{T}$. At high filling,
$S>0$ means that $\mathbf{E}$ is directed parallel to $\nabla{T}$.
This can be interpreted as a change in charge transport from electrons
to holes.  We remark that this sign reversal also occurs in the
insulating as well as in the critical regime.

\begin{figure}
\hspace*{0.75cm}
\resizebox{0.4\textwidth}{!}{
  \includegraphics{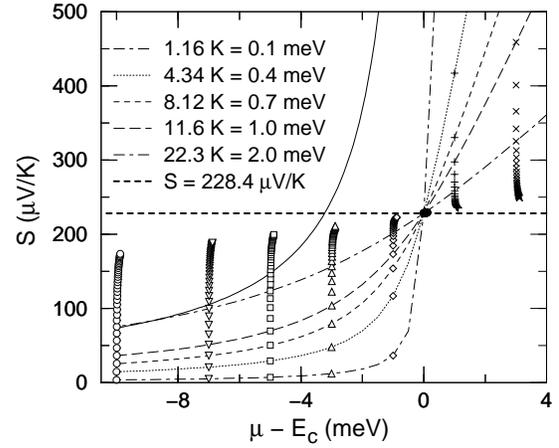}}
\caption{
  The data of $S$ in Fig.\,\protect\ref{tpfig1} shown as a function of
  $\mu$ measured from  $E_c = 7.5$ eV. The horizontal line
  indicates the fixed point MIT value as given in
  Eq.\,(\protect\ref{tpEB}).  The thin dashed lines represent
  isotherms of $S$ calculated using the same method as in
  Ref.\,\protect\cite{enderby}.  The solid line is an isotherm of $S$
  obtained from Eq.\,(\protect\ref{tpCAS}) for $T=22.3$ K.  }
\label{tpfig2}      
\end{figure}

In Fig.\,\ref{tpfig2}, we take the data of Fig.\,\ref{tpfig1} and plot
them as a function of $\mu-E_c$.  Our data coincides with the
isothermal lines which were calculated according to Ref.\ 
\cite{enderby} by numerically integrating $L_{12}$ and $L_{11}$ for a
particular $T$ to get $S$.  We observe that all isotherms of the
insulating ($\mu > E_c$) and the metallic ($\mu < E_c$) regimes cross
at $\mu = E_c$ and $S=228.4\,\mu$V/K.  Comparing with
Eq.\,(\ref{tpCAS}), we again find that the Sommerfeld expansion does
not give the correct behavior of $S$ in the critical regime.

\begin{figure}
\hspace*{0.75cm}
\resizebox{0.4\textwidth}{!}{%
  \includegraphics{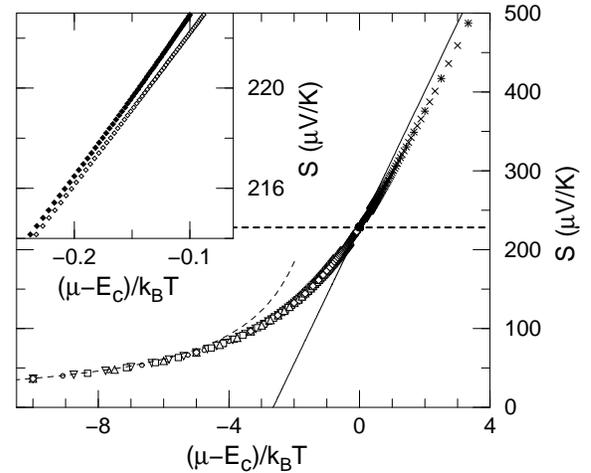}
}                 
\caption{Scaling plot of the thermopower $S$. The thick dashed line
  indicates the fixed point value at the MIT, the solid line
  represents the high-$T$ expansion (\protect\ref{eq-ht-tp}),
  and the thin dashed line shows the Sommerfeld expansion.  The inset
  shows the difference in the scaling when plotting $S$ for $E_F-E_c=
  -0.001$ eV as function of $(\mu - E_c)/k_B T$ (open
  symbols) or $(E_F - E_c)/k_B T$ (filled symbols).  }
\label{tpfig3}
\end{figure}

The data presented in Fig.\,\ref{tpfig2} suggest that one can scale
them onto a single scaling curve. In Fig.\,\ref{tpfig3}, we show that
this is indeed true, when plotting $S$ as a function of $(\mu-E_c)/k_B
T$. We emphasize that the scaling is very good and the small width of
the scaling curve is only due to the size of the symbols. The result for
the high-$T$ expansion is indicated in Fig.\,\ref{tpfig3} by a solid
line. It is good close to the MIT. In the metallic regime, the
Sommerfeld expansion correctly captures the decrease of $S$ for large
negative values of $(\mu -E_c)/k_B T$. We remark that a scaling with
$(E_F-E_c)/k_B T$ as predicted in Ref.\ \cite{sivan} is approximately
valid. The differences are very small as shown in the inset of
Fig.\,\ref{tpfig3}.

\subsection{The Thermal Conductivity and the Lorenz Number}

\begin{figure}
\hspace*{0.75cm}
\resizebox{0.4\textwidth}{!}{
  \includegraphics{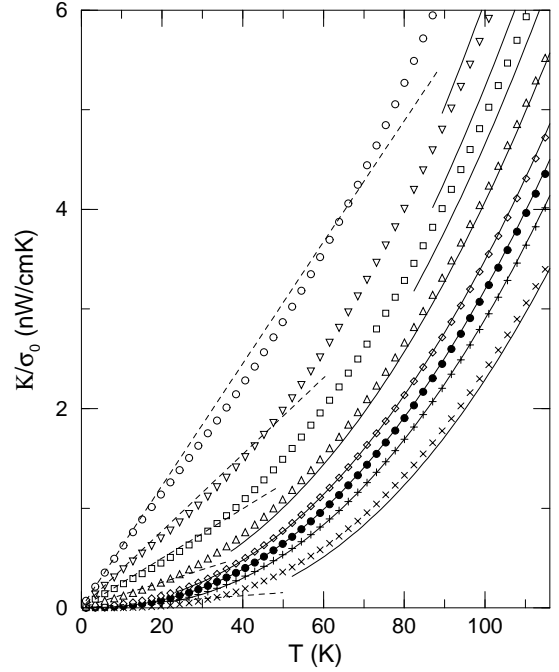}}
\caption{
  The thermal conductivity $K$ as a function of temperature. The
  symbols are as shown in Tab.\ \protect\ref{tab:sym}.
  The dashed lines were obtained in ${\rm O}(T)$ from the Sommerfeld
  expansion (\ref{tc3}) for the metallic regime. The results of the
  high-$T$ expansion for the $8$ choices of $E_F-E_c$ are
  indicated by solid lines.}
\label{thcondfig}       
\end{figure}

In Fig.\,\ref{thcondfig}, we show the $T$ dependence of the thermal
conductivity $K$.  We see that $K\rightarrow{0}$ as $T\rightarrow{0}$
whether it be in the metallic or insulating regime.  We note again
that this simple behavior is due to the fact that our model does not
incorporate phonon contributions. The $T$ dependence of $K$ varies
whether one is in the metallic regime or in the insulating regime and
how far one is from the MIT.  Directly at the MIT, we find that
$K\rightarrow{0}$ as $T^{\nu+1}$ confirming the $T$ dependence of $K$
as given in Eq.\,(\ref{kEB}).  Near the localization MIT, the $T$
dependence of $K/T$ is thus the same as for $\sigma$ in agreement with
Ref.\,\cite{strinati}.  Again, we see that the Sommerfeld expansion
(\ref{tc3}) is reasonable only at low $T$ in the metallic regime. As
for $\sigma$ and $S$, we see that the high-$T$ expansion is
again fairly good in the vicinity of the critical regime.

At this point we are able to determine the behavior of the entropy in
the system as $T\rightarrow{0}$. In the metallic regime, $S$ and $K$
vanish as $T\rightarrow{0}$, while in the critical and insulating
regime, $\sigma$ and $K$ vanish as $T\rightarrow{0}$.  Applying these
results to Eqs.\,(\ref{ecurrent2}) and (\ref{entropy}) yields that for
all regimes the entropy current density $\langle\mathbf{j_q}\rangle/T$
vanishes as $T\rightarrow{0}$.  Therefore, we find that the third law
of thermodynamics is satisfied for our numerical results of the 3D
Anderson model.

\begin{figure}
  \hspace*{0.75cm} \resizebox{0.4\textwidth}{!}{
    \includegraphics{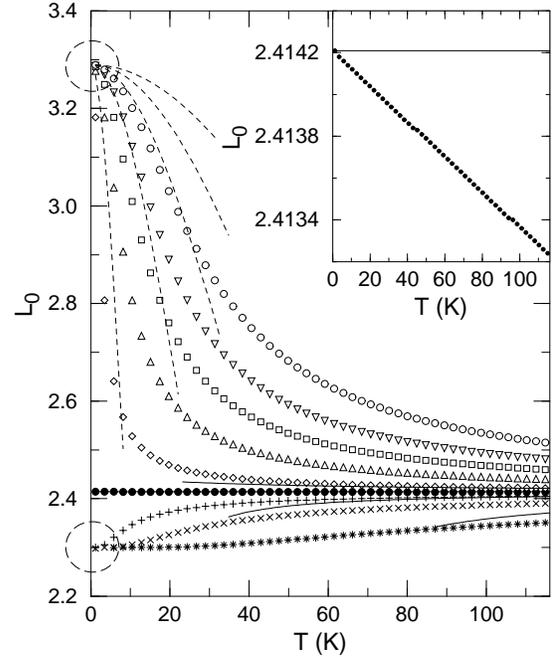}}
\caption{
  The Lorenz number $L_0$ as function of temperature. The
  symbols are as shown in Tab.\ \protect\ref{tab:sym}.
  The dashed circles mark the values of $L_0$ at $T=0$ for metallic
  and insulating regimes. The dashed lines were obtained from Eq.\ 
  (\protect\ref{lo3}). The results of the high-$T$ expansion for
  $E_F-E_c= 0$ eV, $\pm 0.001$ eV and
  $0.003$ eV 
  are indicated by solid lines. The inset shows the
  behavior at $E_F=E_c$ on an enlarged scale.}
\label{Lonumfig}       
\end{figure}

Next, we present the Lorenz number (\ref{lo1}) as a function of $T$ in
Fig.\,\ref{Lonumfig}.  In the metallic regime, we obtain the universal
value $\pi^2/3$ as $T\rightarrow{0}$.  Note that for a metal this
value should hold up to room $T$ \cite{ashcroft}.  However,
our results for the Anderson model show a nontrivial $T$ dependence.
One might have hoped that the higher-order terms in Eq.\,(\ref{lo3})
could adequately reflect the $T$ dependence of our $L_0$ data.
However, this is not the case as shown in Fig.\,\ref{Lonumfig}. This
indicates that even if we incorporate higher order $T$ corrections the
Sommerfeld expansion will not give the right behavior of $L_0$ near
the MIT. We emphasize that the radius of convergence of
Eq.\,(\ref{lo3}) is even smaller than for $\sigma$, $S$ and $K$.
Similarly, the high-$T$ expansion is also much worse than previously
for $\sigma$, $S$ and $K$. Thus in addition to the results for the
critical regime, we only show in Fig.\,\ref{Lonumfig} the results for
nearby data sets in the insulating and metallic regimes.
The $T$ dependence of $L_0$ is linear as shown in the inset of
Fig.\,\ref{Lonumfig}. As before for $S$, the high-$T$
expansion does not reproduce this. At the MIT, $L_{0}=2.4142$.  This
is again the predicted \cite{enderby} $\nu$-dependent value as given
in Eq.\,(\ref{loEB}).

In the insulating regime, one can show analytically by taking the
appropriate limits that $L_0$ approaches $\nu +1$ as $T\rightarrow 0$.
In agreement with this, we find that $L_0=2.3$ at $T=0$ in Fig.\ 
\ref{Lonumfig}. At first glance, it may appear surprising that a
transport property in the insulating regime could be determined by a
universal constant of the critical regime such as $\nu$. However, in
the evaluation of the coefficients $L_{ij}$, the derivative of the
Fermi function for any finite $T$ decays exponentially and thus one
will always have a non-zero overlap with the critical regime. In the
evaluation of Eq.\ (\ref{lo2}), this $\nu$ dependence survives in the
limit $T\rightarrow 0$. In real materials, we expect the relevant
high-energy transfer processes to be dominated by other scattering
events and thus $L_0$ should be different. Nevertheless, for the
present model, this $\nu$ dependence holds.

\subsection{Possible Scenarios in the Critical Regime}

The results presented in Sec.\ \ref{tp} for the thermopower at the MIT
show that $S=228.4\, \mu$V/K for $\nu=1.3$. This value is $2$ orders of
magnitude larger than those measured near the MIT
\cite{lauinger,sherwood,lakner}. However, as mentioned in the
introduction, the conductivity exponents found in many experiments are
either close $\nu=0.5$ or to $1$ \cite{nu} and one might hope that
this difference may explain the small experimental value of $S$.
Also, recent numerical studies of the MIT by transfer-matrix methods
together with non-linear finite-size scaling find $\nu=1.57\pm{0.03}$
\cite{slevin}. In Tab.\ \ref{tab:nu} we summarize the values of $S$
and $L_0$ at the MIT for these conductivity exponents.  We see that
all $S$ values still differ by $2$ orders of magnitude from the
experimental results.  Furthermore, we note that our results for $S$
and $L_0$ are independent of the unit of energy. Even if, instead of
$1\,$ eV, we had used $t_{ij}=1\,$ meV, which is appropriate in the doped
semiconductors \cite{nu,stupp,lakner,liu}, we would still obtain the
values as in Tab.\ \ref{tab:nu}.
\begin{table}[b]
\caption{
  The thermopower and the Lorenz number at the MIT for a 3D Anderson
  model evaluated for various $\nu$ at $E_c=7.5$ eV. The values for
  $\nu=0.5$ and $1$ have already been shown in Ref.\ 
  \protect\cite{enderby}.}
\label{tab:nu}       
\begin{center}
\begin{tabular}{r@{}lr@{}lr@{}l}
\hline\noalign{\smallskip}
\multicolumn{2}{c}{$\nu$}  & 
\multicolumn{2}{c}{$S$}  & 
\multicolumn{2}{c}{$L_0$} \\
& & 
\multicolumn{2}{c}{($\mu$V/K) } &
& \\
\noalign{\smallskip}\hline\noalign{\smallskip}
0. & 5  & 163. & 5  & 1. & 7761 \\
1. & 0  & 204. & 5  & 2. & 1721 \\
1. & 3  & 228. & 4  & 2. & 4142 \\
1. & 57 & 249. & 7  & 2. & 6372 \\
\noalign{\smallskip}\hline
\end{tabular}
\end{center}
\end{table}
Thus our numerical results for the thermopower of the Anderson model
at the MIT show a large discrepancy from experimental results. This
may be due to our assumption of the validity of Eq.\ (\ref{dc_cond})
for a large range of energies, or due to the absence of a true
Anderson-type MIT in real materials, or due to problems in the
experiments.

A different scenario for a disorder driven MIT has been proposed by
Mott, who argued that the MIT from the metallic state to the
insulating state is discontinuous \cite{mott1}. Results supporting
such a behavior have been found experimentally \cite{mott2,moebius}.
According to this scenario, $\sigma$ drops from a finite value
$\sigma_{min}$ to zero \cite{mott1} for $T=0$ at the MIT. This minimum
metallic conductivity $\sigma_{min}$ was estimated by Mott to be
\begin{equation}
\sigma_{min}\simeq\frac{1}{a}\frac{e^2}{\hbar} \label{dc_min} 
\end{equation}
where $a$ is some microscopic length of the system such as the inverse
of the Fermi wave number, $a\approx{k}_{F}^{-1}$.  As summarized in
Ref.\,\cite{mott2}, experiments in non-crystalline materials seem to
indicate that $\sigma_{min}>{300}\;\mathrm{\Omega}^{-1}$cm$^{-1}$.
Let us assume the behavior of $\sigma(E)$ close to the MIT to be
\begin{equation}
\sigma(E)=
\left\{ \begin{array}{cc}
\sigma_{min}, & \quad |E| \leq E_{c}, \\
0,            & \quad |E|>E_{c},
\end{array}\right. 
\label{mmc}
\end{equation}
with $\sigma_{min}={300}\;\mathrm{\Omega}^{-1}$cm$^{-1}$.
Using the numerical approach of Sec.\ \ref{method}, we obtain
$S=119.5\;\mu$V/K at the MIT. This value is still rather large and
thus the assumption of a minimum metallic conductivity as in Eq.\ 
(\ref{mmc}) cannot explain the discrepancy from the experimental
results. We remark that the order of magnitude of $S$ is not changed
appreciably, even if we add to the metallic side of Eq.\,(\ref{mmc}) a
term as given in Eq.\,(\ref{dc_cond}) with $\sigma_0$ a few hundred
$\mathrm{\Omega}^{-1}$cm$^{-1}$ and $\nu=1$.

Lastly, we note that the transport properties calculated for $W = 8$
and $14$ do not differ from those obtained for $W = 12$ in both the
metallic and insulating regions provided we are at temperatures $T
\lesssim 100$K.  For $S$ and $L_0$ at the MIT we obtain the same
values as for $W = 12$.  Again we observe that both $S$ and $L_0$
approach these values linearly with $T$, but with different slopes.
Our results show that the higher the disorder strength the smaller the
magnitude of the slope.

\section{Conclusions}
\label{conclude}

In this paper, we investigated the thermoelectric effects in the 3D
Anderson model near the MIT. The $T$ dependence of the transport
properties is determined by $\mu(T)$. We were able to compute $\mu(T)$
by numerically inverting the formula for the number density $n(\mu,T)$
of noninteracting particles.  Using the result for $\mu(T)$, we
calculated the thermoelectric transport properties within the
Chester-Thellung-Kubo-Greenwood formulation of linear response.  As
$T\rightarrow{0}$ in the metallic regime we verified that $\sigma$
remains finite, $S\rightarrow{0}$, $K\rightarrow{0}$ and
$L_{0}\rightarrow\pi^{2}/3$. On the other hand, in the insulating
regime, $S\rightarrow\infty$.  This we attribute to both $\sigma$ and
$K$ going to zero. Thus, it becomes increasingly difficult to achieve
equilibrium and, hence, the system requires
$\mathbf{E}\rightarrow\infty$.  For $L_0$, we obtained a universal
value of $\nu+1$ even in the insulating regime.
Directly at the MIT, the thermoelectric transport properties agree
with those obtained in Ref.\ \cite{enderby}.  Namely, as
$T\rightarrow{0}$, we found $\sigma\sim T^{\nu}$, $K\sim T^{\nu+1}$,
while $L_0\rightarrow{\mbox{const}}$.

The thermopower $S$ also remains nearly constant in the critical
regime and, in particular, it does not diverge at the MIT in contrast
to earlier calculations using the Sommerfeld expansion at low $T$
\cite{castellani}.  Here we showed that the difference is not so much
due to an order of limits problem, but rather reflects the breakdown
of convergence of the Sommerfeld expansion at the MIT \cite{sivan}.
Our result is supported by scaling data for $S$ at different values of
$T$ and $E_F$ onto a single curve which is continuous across the
transition.  Some of the experiments \cite{lauinger,sherwood} for $S$
have been influenced by the Sommerfeld expansion such that the authors
plot their results as $S/T$. We remark that in such a plot the
signature of the MIT is hard to identify, since $S/T$ at the MIT
diverges as $T\rightarrow 0$ solely due to the decrease in $T$. Our
results suggest that plots as in Figs.\,\ref{tpfig1} and \ref{tpfig2}
should show the MIT more clearly.

The value of $S$ is at least two orders of magnitude larger than
observed in experiments \cite{lauinger,sherwood,lakner}.  This large
discrepancy may be due to the ingredients of our study, namely, we
assumed that a simple power-law behavior of the conductivity
$\sigma(E)$ as in Eq.\ (\ref{dc_cond}) was valid even for $E \ll E_c$
and $E \gg E_c$.  Furthermore, we assumed that it is enough to
consider an averaged density of states $\rho(E)$.  While the first
assumption is of course crucial, the second assumption is of less
importance as we have checked: Local fluctuations in $\rho(E)$ will
lead to fluctuations in the thermoelectric properties for finite $T$,
but do not lead to a different $T\rightarrow 0$ behavior: $S$ remains
finite with values as given in Tab.\ \ref{tab:nu}. Moreover, averaging
over many samples yields a suppression of these fluctuations and a
recovery of the previous behavior for finite $T$.
In this context, we remark that --- naively assuming all other parts
of the derivation are unchanged --- implications of many-particle
interactions such as a reduced single-particle density of states at
$E_F$ \cite{coulombgap}, will only modify the $T$ dependence of $\mu$.
Consequently, the $T$ dependencies of $S$, $\sigma$, $K$, and $L_0$
may be different, but their values at the MIT remain the same.

Our results also suggest that the critical regime is very small.
Namely, as the filling increases slightly from $n= 97.74\%$ to
$97.80\%$, the behavior of the system changes from metallic to
critical and finally to insulating. Up to the best of our knowledge,
such small changes in the electron concentration have not been used in
the  measurements of $S$ as in Refs.\ 
\cite{lauinger,sherwood,lakner}.  We emphasize that such a fine tuning
of $n$ is not essential for measurements of $\sigma$ as is apparent
from Fig.\ \ref{dccondfig}.

Of course, one may also speculate \cite{enderby} that these results
suggest that a true Anderson-type MIT has not yet been observed in the
experiments.

\begin{acknowledgement}
  We thank Frank Milde for programming help and Thomas Vojta for
  useful and stimulating discussions. We gratefully acknowledge
  stimulating communications from John E.~Enderby and Yoseph Imry.
  This work has been supported by the DFG as part of SFB393.
\end{acknowledgement}



\begin{thebibliography}{99}
\bibitem{sommer} The Sommerfeld expansion is used for integrals of the
  form
  $\int^{\infty}_{-\infty}d{\varepsilon}H(\varepsilon)f(\varepsilon)$
  where $f(\varepsilon)$ is the Fermi function and $H(\varepsilon)$ is
  nonsingular and not rapidly varying in a region of order $k_{B}T$
  near $\varepsilon=\mu$.  The integral is expanded as follows
\begin{eqnarray}
\lefteqn{
\int^{\infty}_{-\infty}d{\varepsilon}H(\varepsilon)f(\varepsilon)=
\int^{\mu}_{-\infty}d{\varepsilon}H(\varepsilon) 
}                     \nonumber \\
& &
+\sum^{\infty}_{j=1}(k_{B}T)^{2j}a_{j}
\left.
\frac{d^{2j-1}}{d\varepsilon^{2j-1}}
H(\varepsilon)\right|_{\varepsilon=\mu}
\end{eqnarray}
where $a_j$ are dimensionless numbers.
See Ref.\,\cite{ashcroft} for more details.
\end{thebibliography}
\end{document}